\documentclass[8cm]{article}
\usepackage{amsmath}
\usepackage{graphicx}
\usepackage{authblk}
\usepackage{cite}
\usepackage{caption}
\usepackage{subcaption}
\usepackage[title,titletoc,toc]{appendix}
\title{On the Theory and Algorithm for rigorous discretization in applications of Information Theory}
\author{Venkateshan Kannan, Jesper Tegn\'er *}
\affil{Computational Medicine Unit, Karolinska Institute, Stockholm, Sweden 17176}
\date{}
\begin{document}
\maketitle
\begin{abstract}
We identify fundamental issues with discretization when estimating information-theoretic quantities in the analysis of
data. These difficulties are theoretical in nature and arise with discrete datasets carrying significant implications for the corresponding claims and results. Here we describe the origins of the methodological problems, and provide a clear illustration of their impact with the example of biological network reconstruction.  We propose an algorithm (shared information metric) that corrects for the biases and the resulting improved performance of the algorithm demonstrates the need to take due consideration of this issue in different contexts. 
\end{abstract}

This is investigated in the context  of network inference  \cite{Tegner,Hecker, Markow,Bansal,Hendrick} where information-theoretic methods have been extensively applied \cite{REVEAL,Butte,Reshef,Faith}. Despite its ubiquity, general systematic analysis of the techniques and their underlying assumptions have been few \cite{INem}.Although information entropy, and its variations such as the joint or conditional entropy, is well-defined for discrete variables, its formulation for continuous probability distribution is far from unambiguous. The flexibility in this definition lies at the heart of the issue when applying it consistently to different physical systems. Specifically, entropy estimation in continuous cases requires discretization of the variable values, and the calculation results are not invariant to the choice of discretization. With the focus on reverse-engineering, we describe the fundamental issues with information-theoretic methods, construct examples that highlight them, and propose an algorithm to remedy the situation.   \\

\section{Results}
Mathematically, Shannon entropy of a variable $X$, defined as $H(X) = \sum_{i} -p_i \log (p_i)$ \cite{Shan}, is a well-defined quantity when $X$ is discrete, taking distinct values $i$ with probabilities $p_i$. This is not the case when $X$ is sampled from a continuous probability distribution where an equivalent definition would just give infinity \cite{Lesne}. Nonetheless, there exists a generalization of mutual information (MI) of two variables $X$ and $Y$ with joint probability distribution $P_{X,Y} (u,v)$ and corresponding marginals $P_X (u)$ and $P_Y (v)$  :
\begin{align}
I   &= H(X) + H(Y) -H(X,Y) \nonumber \\
    &=  \sum_{X=u,Y=v} P_{X,Y} (u,v) \log \frac{P_{X,Y} (u,v)}{P_X(u) P_Y(v)}
\end{align}
from discrete to continuous joint probability distribution
\begin{equation}
I= \int du dv P_{X,Y} (u,v) \log{\frac{P_{X,Y} (u,v)}{P_X(u) P_Y(v)}}
\label{MI-C}
\end{equation}
which is well-behaved \cite{Lesne}.  However, given discrete ordered pair ($x_p,y_p$) sampled from an unknown joint probability distribution over continuous variables, we would have to approximate the true (continuous) distribution from it. At the level of probability, this is achieved by binning and discretizing the range of values of the variables. \\

\begin{figure}
\centering
\includegraphics[scale=0.30]{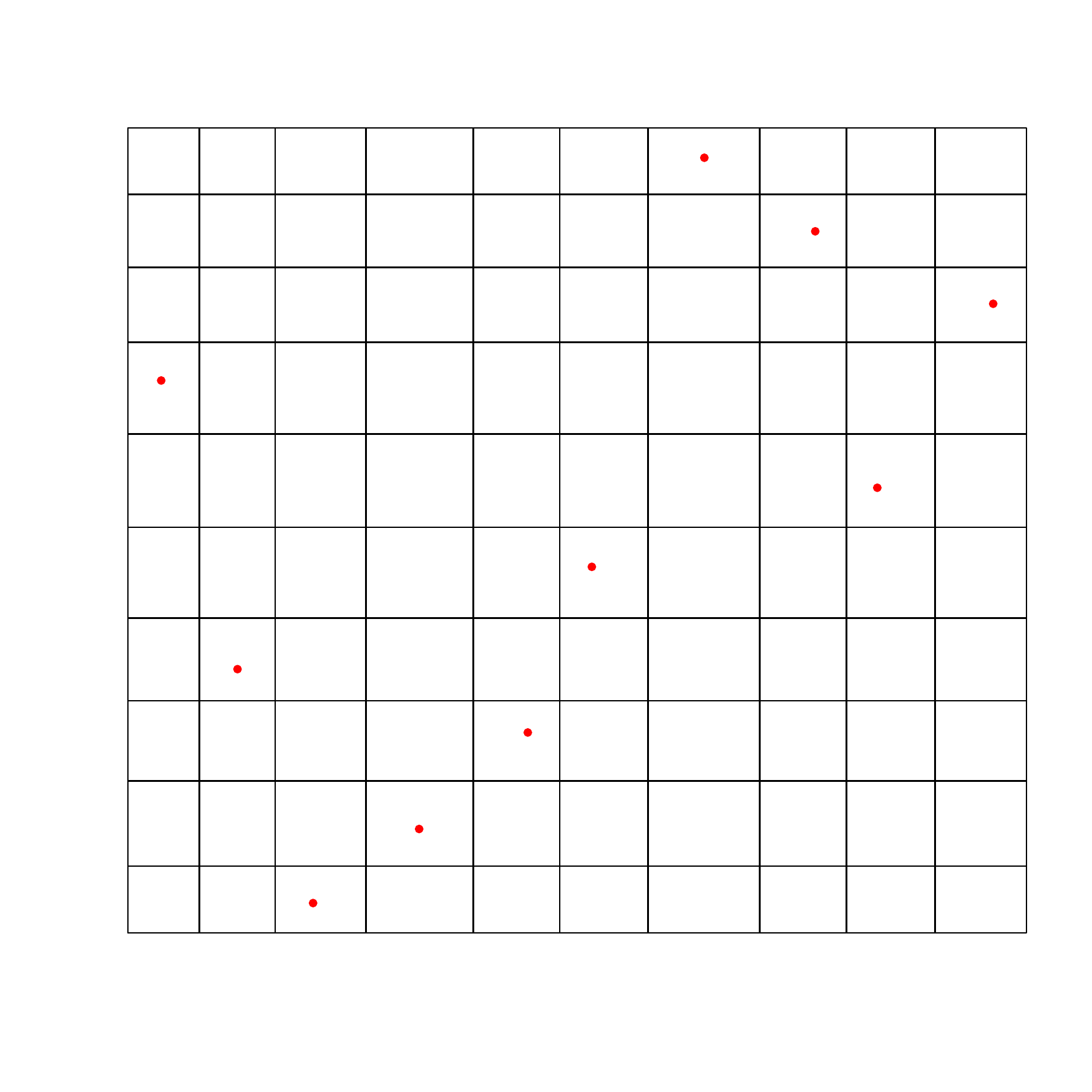}
\caption{\it Grids in 2D space for a set of 10 points such that for any partition along the $X$-direction ($Y$-direction) there is exactly one point within it. This corresponds to a maximum possible mutual information of $\log 10$.}
\label{full-MI}
\end{figure}

While this is carried out straightforwardly, there are pitfalls in simple generalization to mutual information. The key problem is that mutual information depends in a significant way on the discretization parameters, even when the number of points is very large.  We show this with the following simple but general example. Consider ordered pairs $(x_v,y_v) \, , v=1,2,\cdots N$ from a joint distribution with the only requirement that $x_u \neq x_v$ for $u \neq v$ and likewise for $y$'s.  If the entire range of $X$ and $Y$ is considered as a single segment for discretization,  $\Delta^{(0)}_{X}$ and $\Delta^{(0)}_{Y}$ then the joint probability distribution $P_{X,Y} (\Delta^{(0)}_{X},\Delta^{(0)}_{Y})=1$ and hence the marginals are also unity for the corresponding $X$ and $Y$ ranges.  It can be immediately seen that the mutual information is zero. Now, for the other extreme, we choose the interval widths $\delta_X$ and $\delta_Y$ such that there is at most one point lying within each interval (for both the variables). We can always do this by selecting $\delta_X < \min_{u \neq v, u,v =1,2,\cdots ,N} \{ |x_u -x_v|\}$ and $\delta_Y < \displaystyle \min_{u \neq v, u,v =1,2,\cdots ,N} \{ |y_u -y_v|\}$. In that case, the discretized joint probability in each rectangular cell is $1/N$ if it is occupied and 0 otherwise.  In much the same way the marginal probability in each interval is either $1/N$ or 0 for both $X$ and $Y$. Fig (\ref{full-MI}) shows how would be done in the general case (note that interval widths there are not uniform).

 We have then the mutual information:
\begin{equation}
I= \sum_{\alpha,\beta} P_{X,Y}(\alpha,\beta) \log \frac{P_{X,Y} (\alpha,\beta)}{P(\alpha) P(\beta)} =\sum_{\alpha,\beta}  1/N \log \frac{1/N}{(1/N)(1/N)} = \log N
\end{equation} 
Thus, for a set of $N$ observations from a joint probability distribution, the mutual information can vary between 0 and $\log{N}$ depending on the size of the binning. What is most striking about this is that {\it this is true regardless of the true mutual information of the underlying joint probability distribution}. As the upper bound grows with the size $N$, having a larger sample number does not solve the fundamental problem of the inherent ambiguity of defining mutual information using discrete data points. \\
\subsection{Standard partitioning biases}
Recognizing this basic limitation, we wish to consider the implications of different choices of binning in estimating mutual information \cite{Simoes}. We want to understand the underlying biases and the possible deviation of mathematical properties from that of the original definition. \\% We want to understand the underlying biases better evaluate methods based on how well they reproduce the relative rankings of their true mutual information \cite{MRNET} and not their absolute value \cite{Altay}. 

\begin{figure}
\begin{subfigure}{0.45 \textwidth}
\includegraphics[scale=0.32,type=pdf, ext=.pdf,read=.pdf]{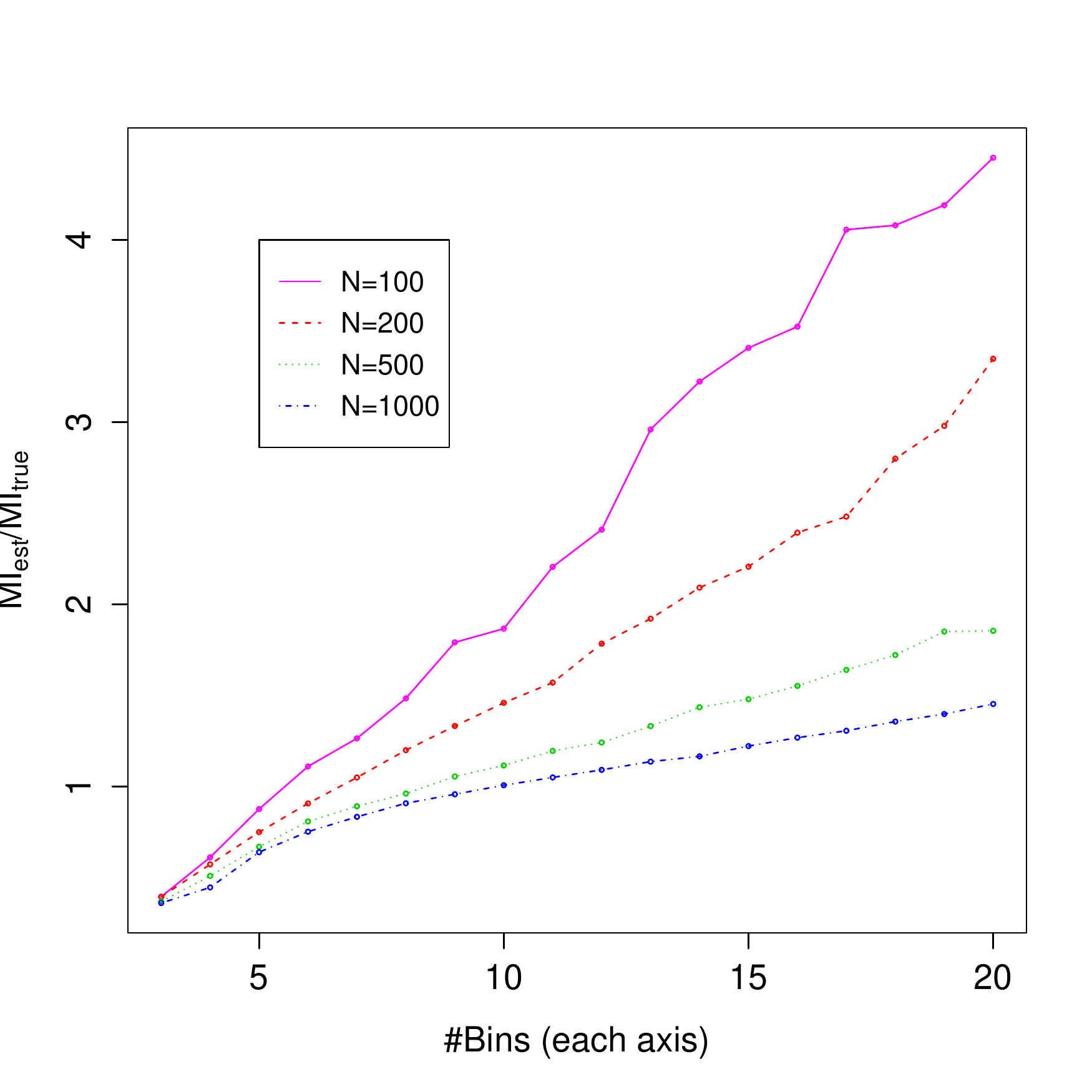}
\caption{(Theoretical) I=0.22}
\end{subfigure}
\begin{subfigure}{0.45 \textwidth}
\includegraphics[scale=0.32,type=pdf, ext=.pdf,read=.pdf]{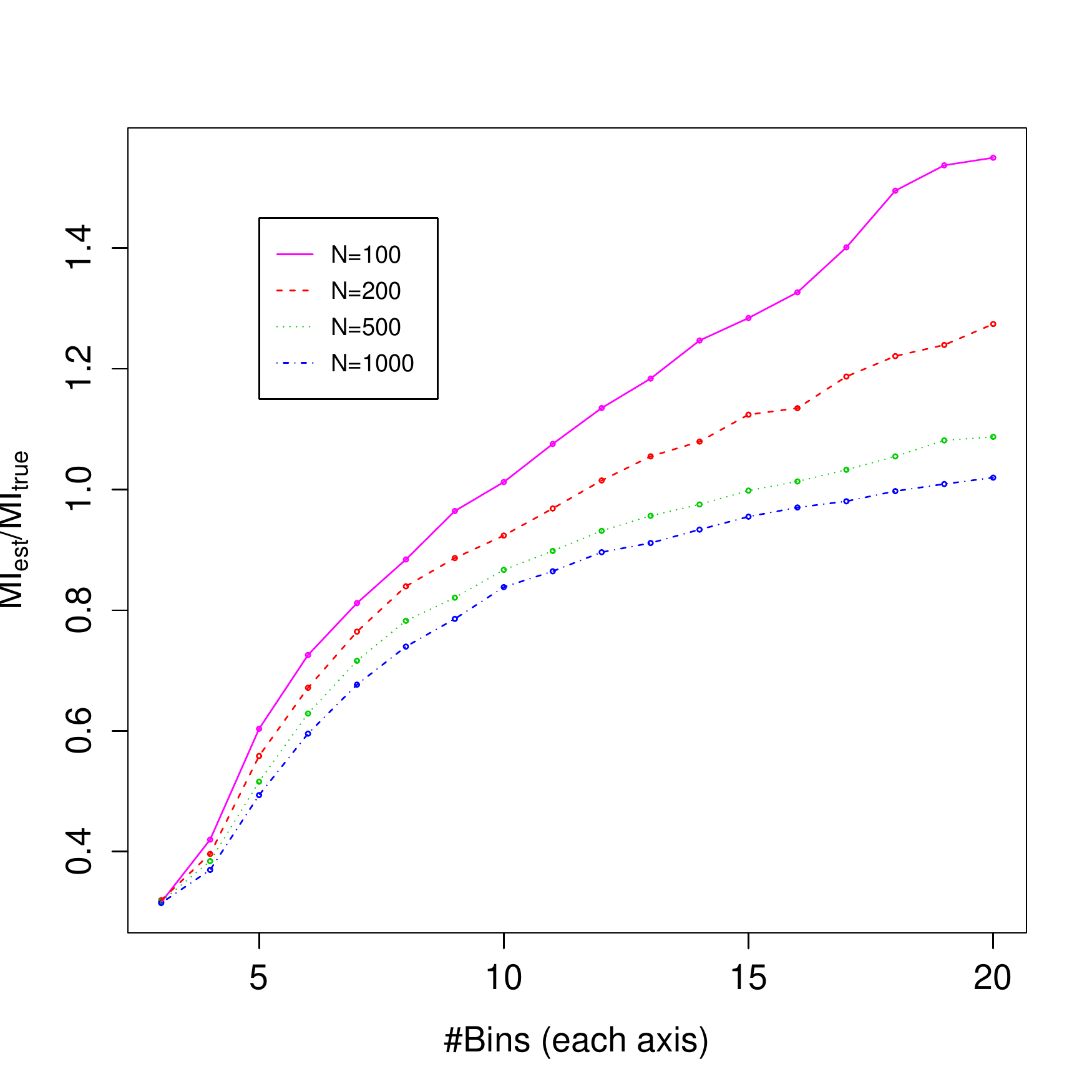}
\caption{(Theoretical) I=0.8}
\end{subfigure}
\caption{\it Variation of the estimated entropy as a function of number of bins (along each axis) for a pair of jointly Gaussian random variables.}
\label{MI-V-Bin-1}
\end{figure}

In general, mutual information increases as we increase the number of partitions of the space.  We can see that in Fig (\ref{MI-V-Bin-1}) the variation of the estimated mutual information for two multivariate Gaussian random variables with the number of bins (assuming fixed bin length)  for different sample sizes $N$. The plotted values for the estimated MI comes from the average taken over 50 samples. In the supplement, we prove a general result stating that the doubling of partition size would increase the mutual information between two variables for an arbitrary set of samples. However, this by itself would not be an issue if we are considering reverse engineering in networks as long as the rankings of estimated mutual information between variables is preserved with respect to their true values \\

There are other biases and errors that are introduced by the choice of the partitioning scheme that we will consider here. Broadly, partitioning of the two dimensional space of variables $X$ and $Y$ fall into two general categories: uniform number of bins ($b$) where the number of divisions of both axis is identical, uniform width where the width of each partition ($w$) is identical for both variables, and uniform frequency where the number of points falling into each partition (for each variable) variable is fixed.. We will now turn to each of these specific methods.  

\subsubsection{Uniform number of partitions} 
\label{Uni-Num}
Here each axis of the space is split into equal number of uniform segments. The advantage of this method is that it works well when the data is distributed reasonably evenly across the range of each axis which happens when the density of points falls off sharply outside of the confined region under consideration. \\

However, the difficulty arises with probability densities that have a fat tail, where the spread is wide but the majority of points are still lying within a smaller region. A binning of this sort would then be finer for points in the periphery but coarser in the denser regions, thus tending to underestimate the mutual information from the latter areas. 

\begin{figure}
\begin{subfigure}{0.43 \textwidth}
\includegraphics[scale=0.5,type=png,ext=.png,read=.png]{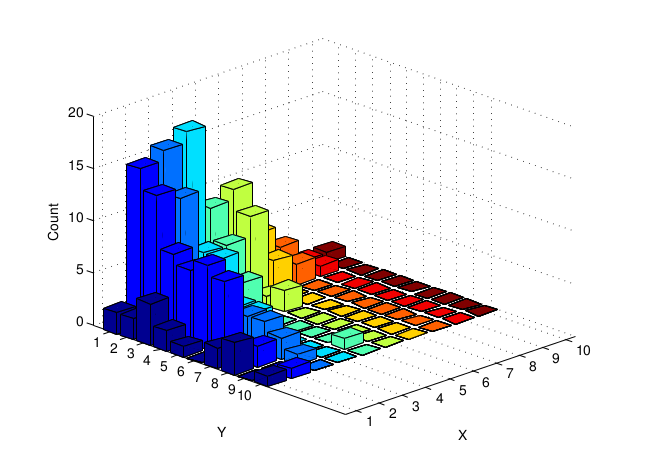}
\caption{\it \small Sample from an exponential family Eq. (\ref{exp}) with $\theta=3$ and $\lambda=5$. Theoretical (estimated) mutual information is 0.79 (0.21) }
\label{Id-NP}
\end{subfigure}
\hspace{1cm} 
\begin{subfigure}{0.43 \textwidth}
\includegraphics[scale=0.5,type=png,ext=.png,read=.png]{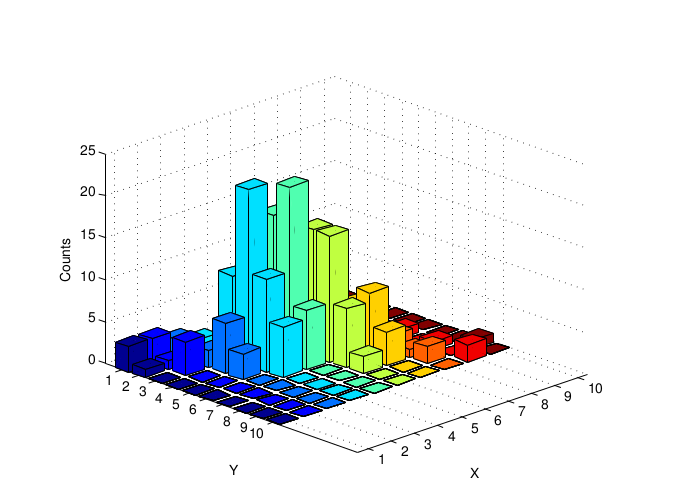}
\caption{\it \small Normal distribution with $\sigma_{xx}^2 = \sigma_{yy}^2=5, \sigma_{xy} = 4$.  Theoretical (estimated) mutual information is 0.51 (0.59)  }
\label{NId-NP}
\end{subfigure}
\newline
\begin{subfigure}{0.43 \textwidth}
\includegraphics[scale=0.4,type=png,ext=.png,read=.png]{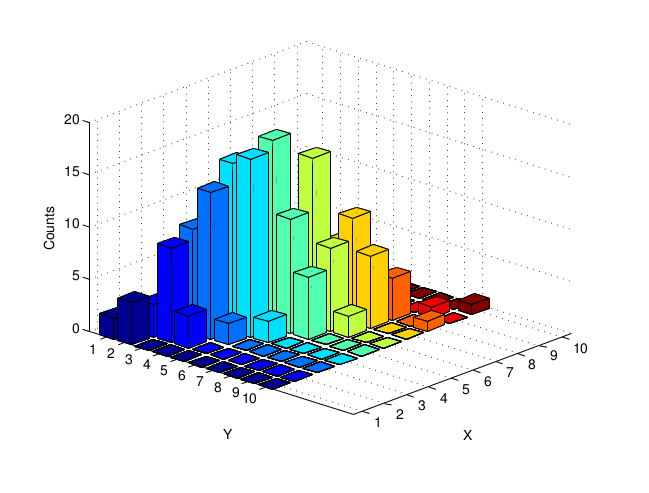}
\caption{\it  Normal distribution with $\sigma_{xx}^2 =5,  \sigma_{yy}^2=5, \sigma_{xy} = 4.5$.  Theoretical (estimated) mutual information is 0.83 (0.79) }
\label{G1}
\end{subfigure}
\hspace{1cm}
\begin{subfigure}{0.43 \textwidth}
\includegraphics[scale=0.4,type=png,ext=.png,read=.png]{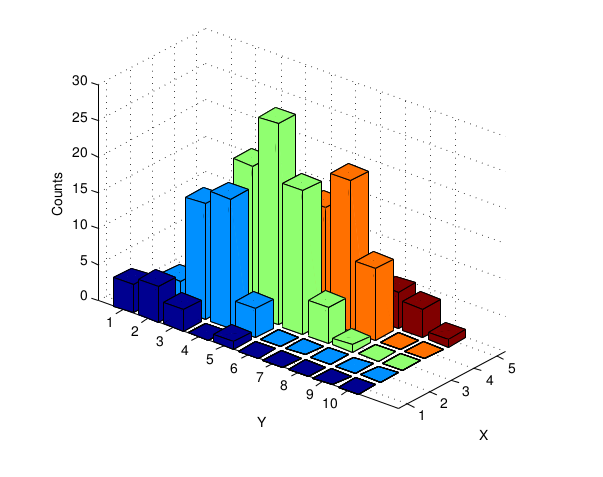}
\caption{\it Normal distribution with $\sigma_{xx}^2 = 5, \sigma_{yy}^2=2.5, \sigma_{xy} = 3.18$.  Theoretical (estimated) mutual information is 0.83 (0.67) }
\label{G2}
\end{subfigure}
\caption{\bf  Mutual Information and meshing:  In (a) and (b), the grid choice turns the space of points into a square with equal number of partitions along both, while for (c) and (d) the width of every partition is held fixed.}
\label{NP}
\end{figure}

We demonstrate this by estimating the entropy for two samples , the first of which comes from the exponential family, 
\begin{equation}
P_{X,Y} = \frac{1}{\Gamma (\theta) \lambda} x^{\theta} e^{-\frac{xy}{\lambda} -x}
\label{exp}
\end{equation}  with $\theta =4, \lambda=5$ and the second a multivariate Gaussian in two-variables $ \sim e^{-(X - \mu_x)\cdot \Sigma^{-1} \cdot (Y - \mu_y)}$ where 
$$\Sigma =  \left( \begin{array}{cc} 5 &4 \\ 4 & 5 \end{array}  \right) $$ 
and $(\mu_x,\mu_y) = (5,5)$. Each set consists of 200 points, and the barplots  of distributions (Fig.(\ref{Id-NP},\ref{NId-NP})) across a uniform $10 \times 10$ grid show that while the multivariate normal is spread quite smoothly across the entire surface, it falls off in the other for $X,Y >5$.

 The mutual information can be theoretically computed in these two cases:
$$ I^{exp} (\theta, \lambda) = \lambda ( \psi (\theta +1) -\log \theta )$$
where $\psi$ is the digamma function and 
$$I^{Gauss} (\Sigma) = \frac{1}{2} \log \frac{\sigma_{xx}^2 \sigma_{yy}^{2}}{\sigma_{xx}^2 \sigma_{yy}^2 -\sigma_{xy}^2}$$
which evaluate to $I^{exp} = 0.79$ and $I^{Gauss} = 0.51$ but the estimated mutual information from the above binning leads to $I^{exp}_e =0.21$ and  $I^{Gauss}_e=0.59$. \\

The reason for the severe underestimation in the first case  is precisely the fact that there is a closer clustering of points within a smaller region where this type of binning tends to be `too wide' and averaging the finer distinctions. In the multivariate normal case, this difference is not so striking, and the estimation is far more accurate.

%For a pair of variables that are jointly Gaussian, there is a closed form expression for the mutual information
%\begin{equation}
%I^{(true)} = \frac{1}{2} \log \frac{\sigma_{xx}^2 \sigma_{yy}^{2}}{\sigma_{xx}^2 \sigma_{yy}^2 -\sigma_{xy}^2}
%\label{TMI}
%\end{equation} 

\vspace{6mm}
\begin{figure}
\begin{subfigure}{0.43 \textwidth}
\includegraphics[scale=0.5,type=png,ext=.png,read=.png]{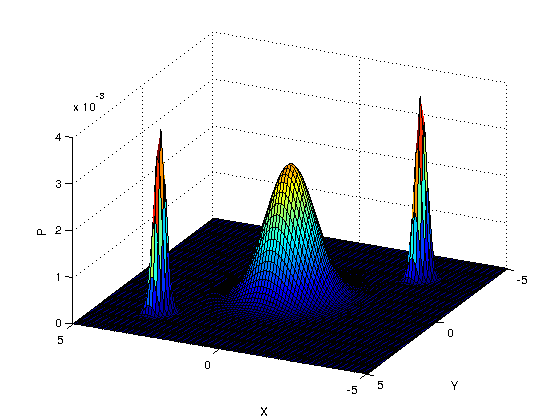}
\caption{ }
\end{subfigure}
\hspace{1cm} 
\begin{subfigure}{0.43 \textwidth}
\includegraphics[scale=0.5,type=png,ext=.png,read=.png]{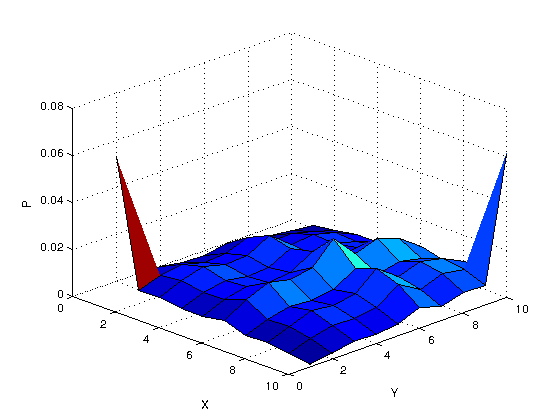}
\caption{ } 
\end{subfigure}
\caption{ (Left) {\it \small Mixed-Gaussian distribution. } (Right){\it \small Equal-frequency partitioning skews the distribution. Estimated(Numerical) mutual information is 0.19 (0.45)} }
\label{EF}
\end{figure}

\subsubsection{Uniform partition widths} 
\label{Uni-Size}
The other standard approach to choosing a grid on the two dimensional space of points is to set the width of every partition on both axes to be a constant. As the spread of the two variables is in general different, the total number of partitions need not be the same. The idea behind this is that the width represents the distance over which points are considered to fall within the same discrete category.  It may be set based on estimates of noise in the sample, or, where available, prior knowledge of the form of the joint distribution. \\

There is however a fundamental problem with this approach : setting the widths to be a constant destroys the scaling invariance of the theoretical definition of mutual information for continuous variables.  In Eq. (\ref{MI-C}), the integral is invariant under the transformation  $y \rightarrow cy, P_{X,Y} (x,y) \rightarrow c P_{X,Y} (x,cy), P_Y (y) \rightarrow c P_Y (cy)$ which corresponds to scaling the metric of the Y-axis by a constant. It must be noted that this is not an incidental property of the integral but one that is central to the usefulness of mutual information, i.e., mere scaling of the underlying space should not, and does not,  alter the relative information between two variables. \\

Choosing binning styles with equal widths implies that scaling the points along one direction in the space would change the number of partitions on its axis, which changes the estimated mutual information.  The effect of this can be easily observed by considering samplings from two multivariate normal distributions in two dimensions that differ only by the scaling of the y-axis by a factor of $1/\sqrt{2}$( Fig. (\ref{G1}) and Fig. (\ref{G2}).  By definition, the scale-invariance implies equal mutual information but the requirement of setting equal widths (unity in this case) leads fewer bins in the second case and consequent underestimation of its value.

%If partition widths $w_i$ are held fixed, the number of partitions would depend on the spread of the node variables, with larger divisions for greater spread and vice-versa.  As it has been argued above that mutual information increases, for the same distribution of points, with increase in the number of partitions, it follows that there is a likely positive bias node variables whose spreads are greater than those that are smaller. \\
\subsubsection{Equal Frequency Partitioning}
In this case, the boundaries of the partitions for a given variable are chosen such that each division contains an equal number of points.  The general advantage of this method is that, unlike the above two, every row or column in the grid would contain equal number of points, eliminating the problem highlighted in Fig (\ref{Id-NP},\ref{NId-NP}). However, this would skew joint distributions that have strong association between two variables to a more uniform shape, causing an underestimation in mutual information. We can see this with the mixed-Gaussian distribution with three modes (Fig. (\ref{EF})) where the effect of choosing the equal frequency partitioning on a $10 \times 10$ grid flattens the peak in the center spreading the probability over a wider region, leading to a decrease in mutual information.  \\

It is thus clear from the examples above that the standard types of partitioning is in general unsatisfactory in terms of obtaining reasonable and consistent mutual information scores. It should be noted that although our examples consider discrete binning of the data, the kernel-based estimation of mutual information \cite{KBE,ARACNE} has the same fundamental problem. While it is not as pronounced as it is with direct binning, the width of the kernel in that case is equivalent to the width of the bins here (equal-width partition).  The same is true with estimators like the k-nearest neighbors \cite{Knn} where the scores depend on the parameter corresponding to the number of neighbors.
\subsection{Adaptive Partitioning for Network Inference}
Data-driven reverse-engineering of genetic networks uses pairwise association between genes to determine true interactions among them.  Information theory is applied effectively for this task by estimating the mutual information between every pair of network node variables and reconstructing the network based on the relative rankings of the different edge scores \cite{Chow,Reshef,Faith}. A distinct advantage being that mutual information detects all forms of associations while correlation-based measures perform well only with linear relations. \\

We propose a novel method of partitioning of space in the context of network reverse-engineering that aims to reduce the biases created by the standard techniques. This includes choosing grid sizes that take into consideration (a) overall spread of the data among all variables (b) spread of the values of the pair of node variables in question (c) dependence of numerical estimation of mutual information on this spread (d) appropriate normalization using the entropy of each variable. \\

Given set $V$ of N node variables and M samples, we choose first a standard partition width $w_0 = w_{int}/K_{min}$, where $w_{int} = \text{median} \{ \sigma (X) | X \in V  \}$, and $K_{min}$ is an integer parameter that represents the minimum number of partitions and $\sigma (X)$ is the empirical standard deviation of the $X$ variable values. \\

For any two node variables $X$ and $Y$ whose mutual information we want to estimate, the number of bins $b_X$ and $b_Y$ are obtained using the following algorithm (see Fig. \ref{alg}). In addition to $K_{min}$, we have another integer parameter $K_{max}$ that sets the maximum number of partitions.  \\
\begin{figure}[t!]
\centering
\framebox{ \begin{minipage}{3in} 
\begin{small}
\hfill \\
{\bf Step 1}: $b_X$ is first set to $[ \frac{\sigma(X)}{w_0} ]$ where $[.]$ refers to the nearest integer function. Likewise for $b_Y$. Assume $b_X \leq b_Y$. \\
{\bf Step 2} : If $b_Y > K_{max}$, then reset $b_X = [ b_X (K/b_Y) ]$ and $b_Y= K$.   If the new $b_X < K_{min}$, reset $b_X=K_{min}$. \\
{\bf Step 3}: If $b_X < K_{min}$, then $b_Y = [ \max \{K_{min}, \min \{ K_{max}, K_{min} \sqrt{b_Y/b_X} \} \} ]$ and $b_X = K_{min}$. \\
{\bf Step 4}: Once the binning is fixed, we proceed to calculate the number of points falling within each rectangle, and the discretized form of mutual information between the two variables. \\
{\bf Step 5}: We normalize the mutual information by dividing by $\min \{ H(X), H(Y) \}$, where $H(X)$ and $H(Y)$ are the entropies of $X$ and $Y$ calculated using the same bin numbers $b_X$ and $b_Y$ respectively.  We call the resulting quantity shared information metric (SIM). \\
\vspace{1cm}
\end{small}
\end{minipage}
 } \\
\caption{ \bf Adaptive Paritioning Algorithm} 
\label{alg}
\end{figure}
\vspace{1cm}

Steps 1-3 may look complicated but the idea is simple enough: we prefer to maintain the bin widths constant as long as their numbers lie between $K_{min}$ and $K_{max}$. This is primarily in recognition of the fact that spread of the node variable is a measure of the strength of the interactions: if the values are nearly the same, the uniform width approach would select fewer partitions. However, to ensure that the distribution is not skewed when the spread of one or both is large (or too small), we need to correspondingly {\it rescale both variables} to force them to lie within that range. Heuristically, the two limits $K_{min}$ and $K_{max}$  ensure that the relative values of mutual information are not under- or over-estimated.  \\

{\bf Normalization Measure}: \\
Step 5 was motivated by two considerations. First, we note that a proper normalization is required when having different bin sizes. If, for example, there is a node variable with a small spread but happens to have a strong direct interaction with another node, the binning scheme would then `overcorrect' for the small spread. We overcome that by such a normalization scheme. Second, we recognize that what is significant in reverse-engineering is not the absolute value of mutual-information but the MI relative to the information content of the two node variables. The inequality $I (X,Y) \leq \min \{ H(X), H(Y) \} $ captures the fact that this quantity represents that part of the information that is shared between the nodes. The fraction of what is shared should serve as a better indicator of true interaction than their absolute values.  \\ 
\subsection{Evaluation on networks}
\begin{figure}
\begin{subfigure}{0.45 \textwidth}
\includegraphics[scale=0.3,type=pdf,ext=.pdf,read=.pdf]{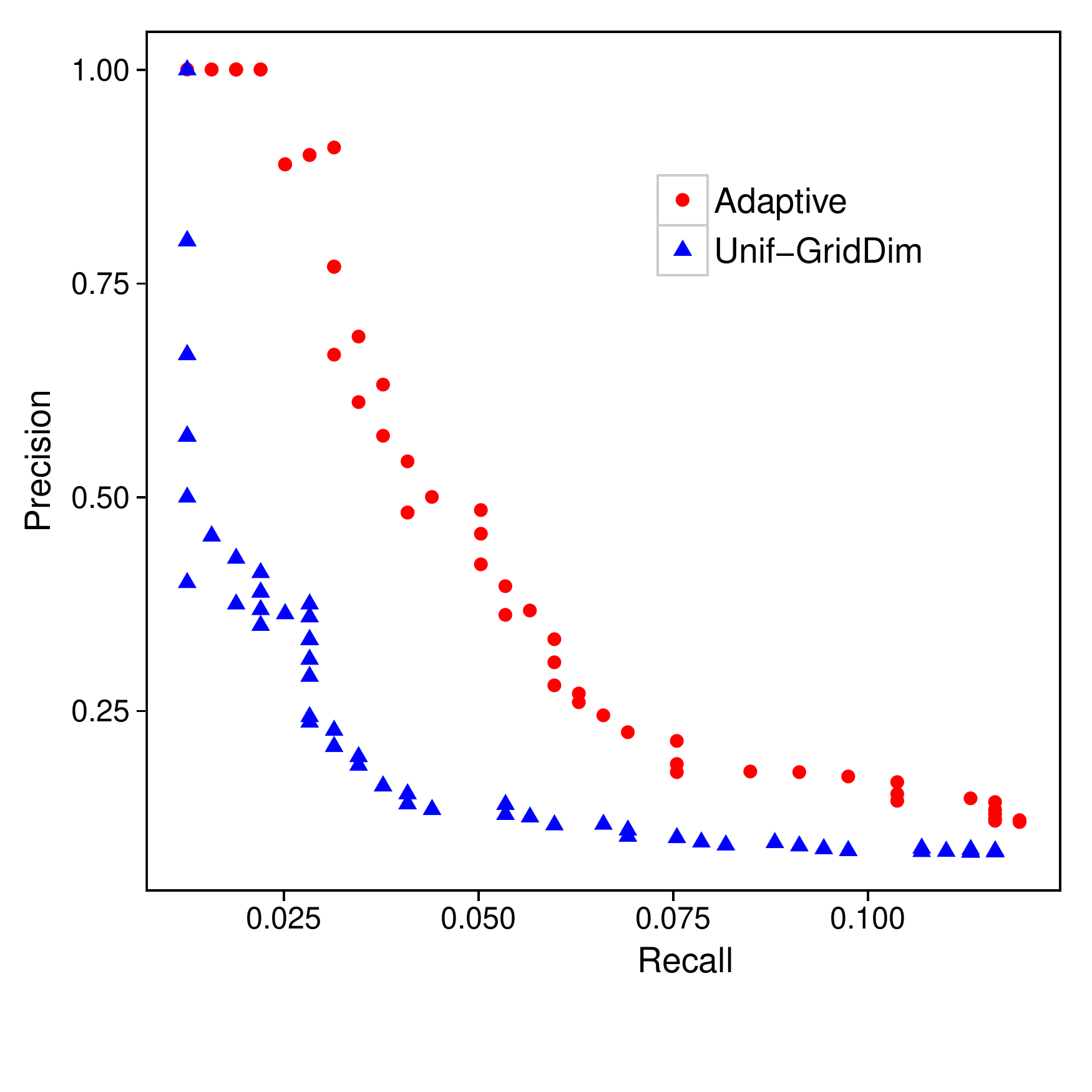}
\caption{$K_{min}=4$ and $K_{max}=10$. $N$ = 100 and $E$= 160}
\end{subfigure}
\begin{subfigure}{0.45 \textwidth}
\includegraphics[scale=0.3,type=pdf,ext=.pdf,read=.pdf]{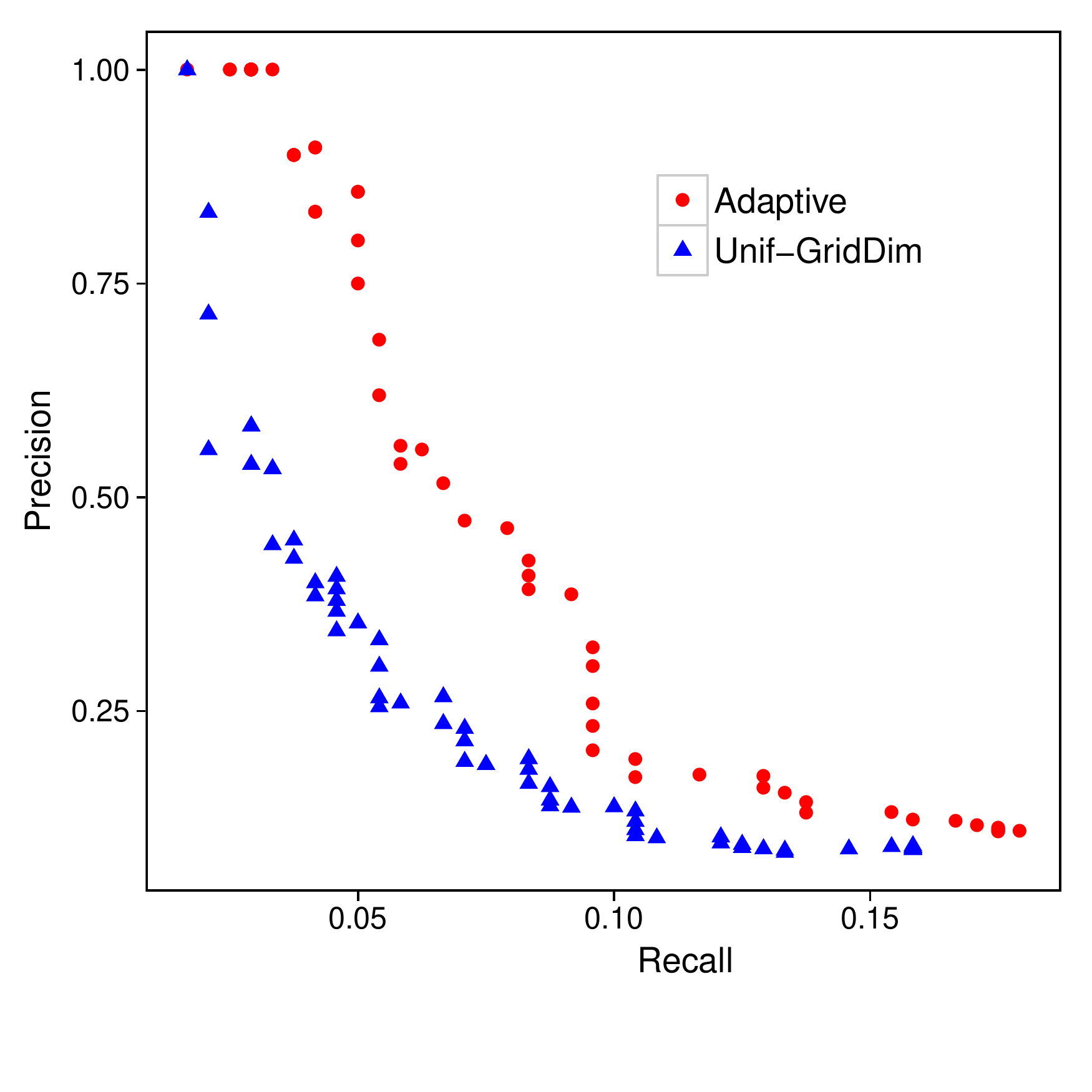}
\caption{$K_{min}$=2 amd $K_{max}$ =5.  $N$ = 100 and $E$= 120}
\end{subfigure}
\newline
\vspace{4mm}
\centering
\begin{subfigure}{0.5 \textwidth}
\includegraphics[scale=0.3,type=pdf,ext=.pdf,read=.pdf]{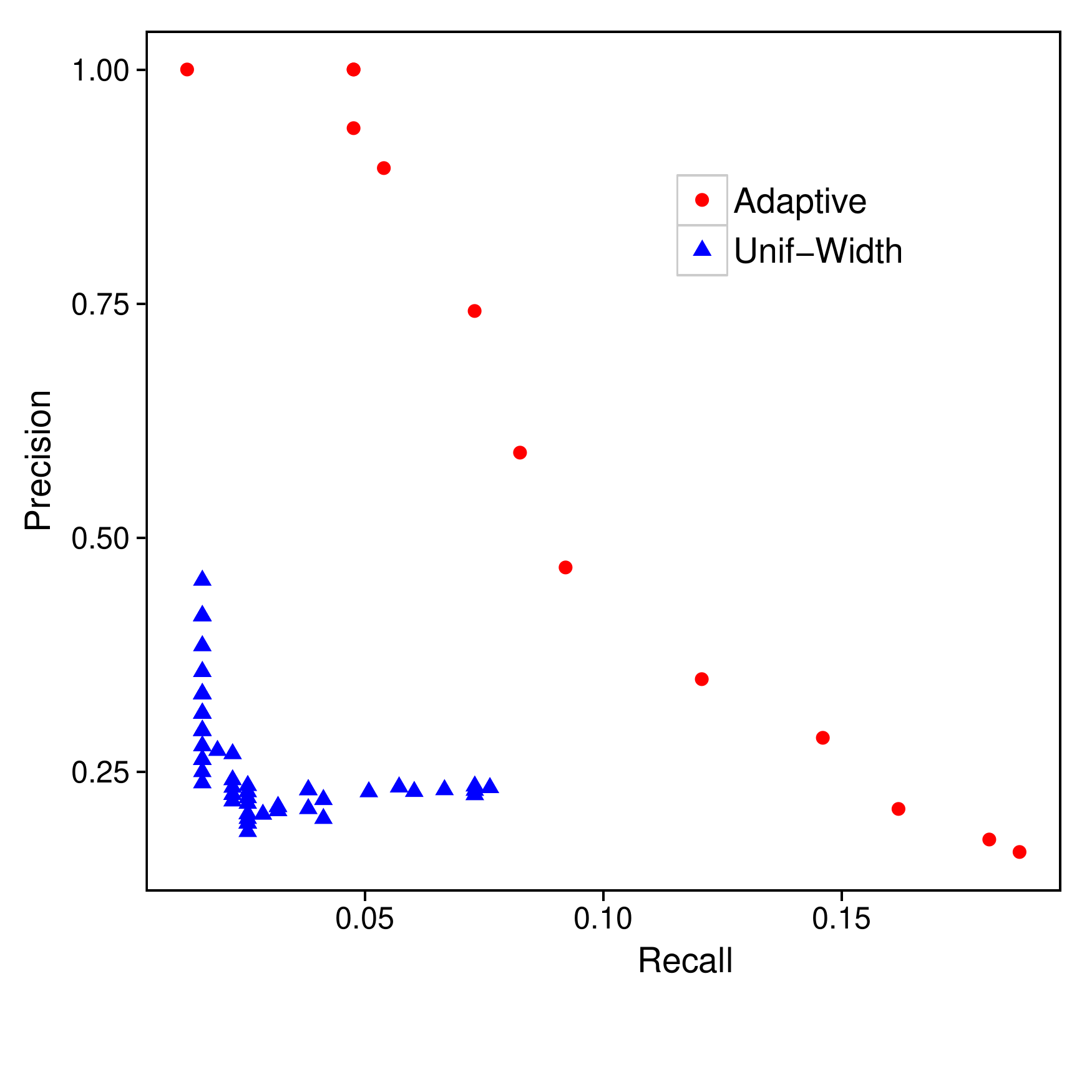}
\caption{Equal-width vs Adaptive Discretization: $N=100$ and $E=160$}
\label{EW-NU}
\end{subfigure}
\caption{\it Comparison of the performance of network edge determination using uniform number off bins (10 and 5 in figure (a) and (b) respectively) or uniform bin width (Fig. (c)) against a non-uniform discretization. $N$ and $E$ refer to the total  number of nodes and true edges in the network.}
\label{PR-UvNU}
\end{figure}

\begin{figure}
\begin{subfigure}{0.45 \textwidth}
\includegraphics[scale=0.3,type=pdf,ext=.pdf,read=.pdf]{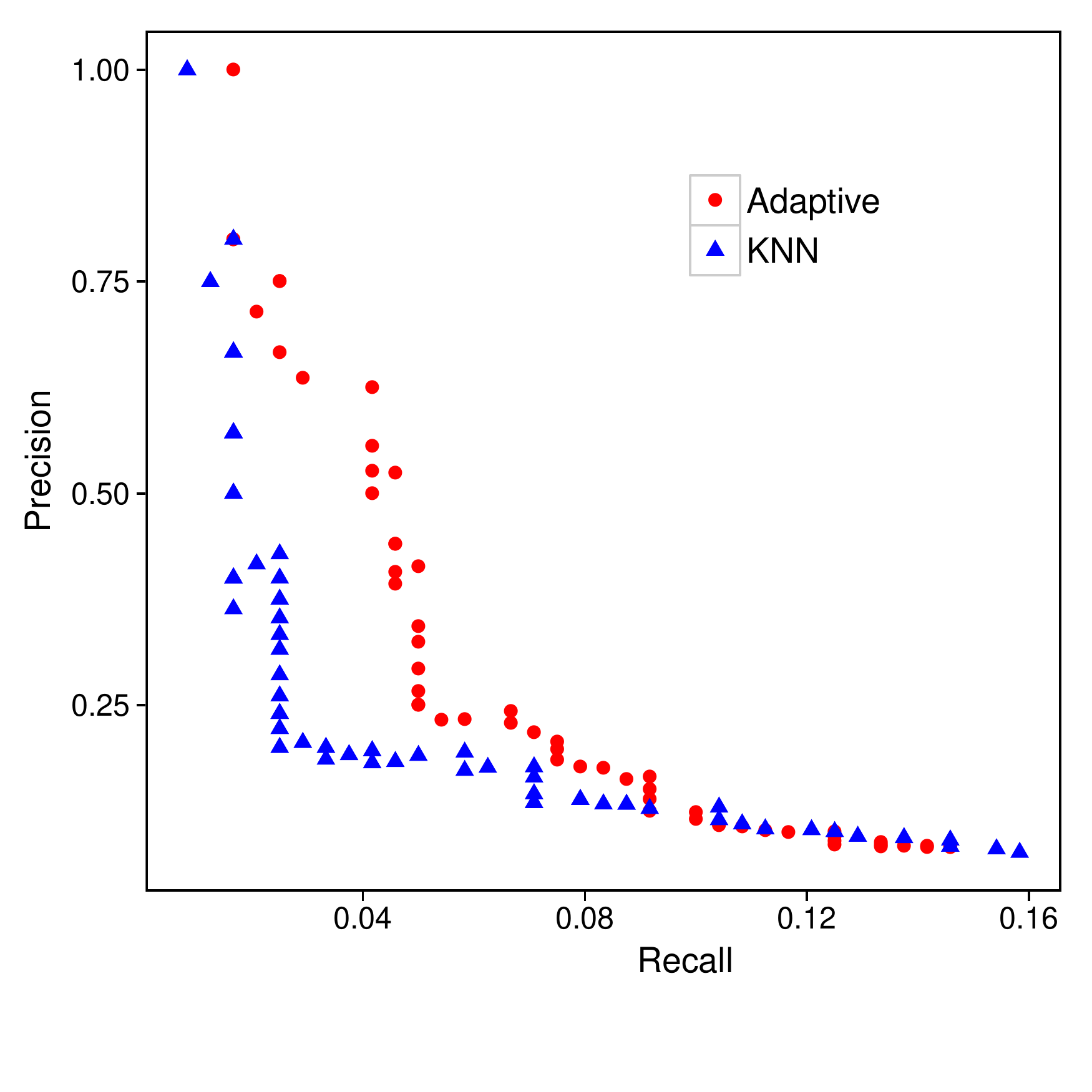}
\caption{\it $N=100$ and $E=120$}
\label{NU-KNN}
\end{subfigure}
\begin{subfigure}{0.45 \textwidth}
\includegraphics[scale=0.3,type=pdf,ext=.pdf,read=.pdf]{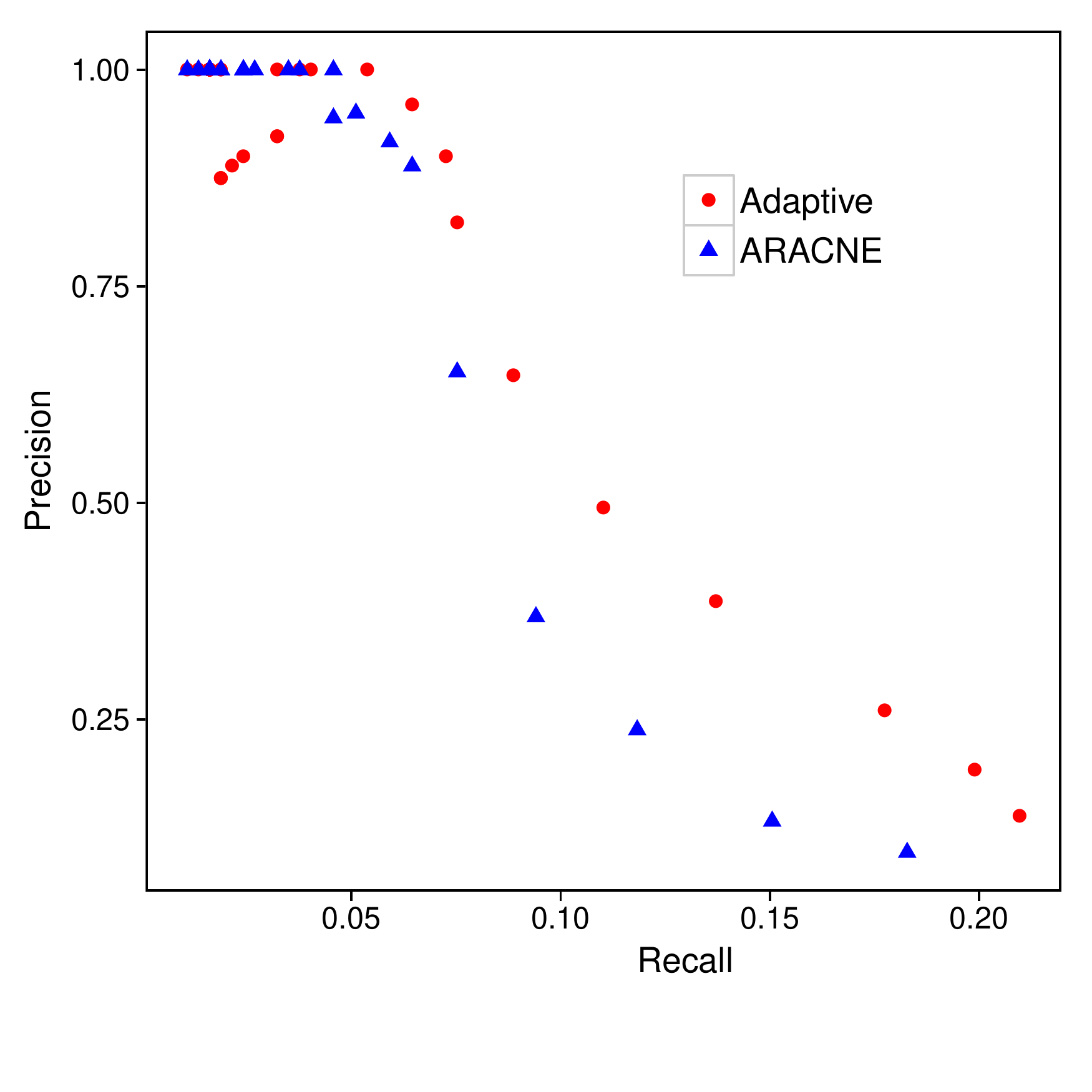}
\caption{\it  $N=100$ and $E=185$}
\label{NU-KNN}
\end{subfigure}
\caption{\it Comparison of the performance of network edge determination using k-nearest neighbor method (with DPI) (Fig. (a)) and ARACNE (Fig (b)) against the non-uniform discretization with the shared information mertic (SIM). $N$ and $E$ refer to the total  number of nodes and true edges in the network.}
\label{KNN-ARACNE}
\end{figure}

\begin{figure}
\centering
\includegraphics[scale=0.3,type=pdf,ext=.pdf,read=.pdf]{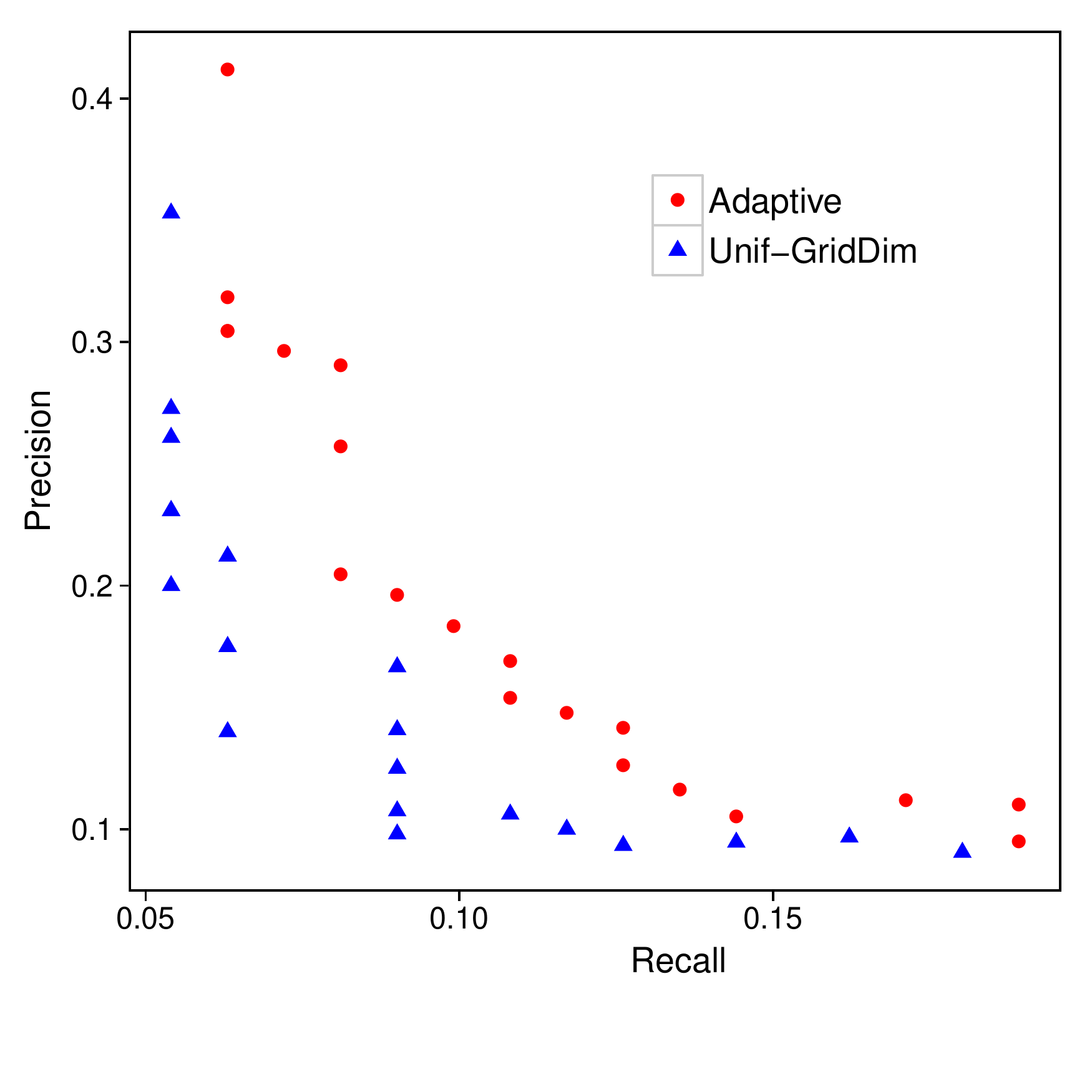}
\caption{\it Comparison of the performance of identification of targets of transcription factors of the gene regulatory network of {\it E. coli}. Expression data and true interaction set obtained from the DREAM 5 challenge.}
\label{EC-C1}
\end{figure}

We assess the algorithm by studying its performance on {\it in-silico} networks against that of standard methods (see Section \ref{Meth} for more details). The initial comparison is with respect to the estimation method using only uniform bin size or uniform bin number. Evaluation of the performance is done by plotting the precision-recall curve (Fig. (\ref{PR-UvNU})) (see Section \ref{Meth} for more details on their definition). It can be clearly seen that, at any given value of recall, the precision is significantly higher when using adaptive non-uniform discretization. We also consistently found better results regardless of the size or the topology of the network or its dynamics.  \\
\\
Our method performed better even against standard mutual-information based  approaches such as kernel estimation and DPI (ARACNE) \cite{ARACNE} and the k-nearest neighbor algorithm (with DPI) \cite{Knn} (Fig. (\ref{KNN-ARACNE})).   
\\
For biological data sets, we considered the gene regulatory network of {\it E. coli}, and the expression data for it from the DREAM 5 challenge \cite{Marbach}.  In this case, the set of transcription factors is known, and the aim is to determine their targets. Thus our binning choice should reflect this prior knowledge and the algorithm was modified accordingly (described in the supplement). Of the set of 2000 interactions we evaluated, of which only 111 represented true regulatory interactions, our method clearly performed better than the standard method of discretization (Fig. \ref{EC-C1}).

\section{Methods}
\label{Meth}
We constructed synthetic networks obeying power-law degree-distribution with the R package {\it igraph}.  The data was generated by carrying out numerical integration of the Hill-kinetic equations corresponding to the network topology until global steady state was reached. Each sample in the data set corresponds to the steady state value of all variables. Different samples correspond to steady states of different equations obtained with set of parameters comes from a uniform distribution. \\

The precision-recall curve (PRC) is used to evaluate these networks, where precision and recall are given by:
$$ \text{Precision} = \frac{TP}{TP + FP}$$ \\$$ \text{Recall}=\frac{TP}{TP +FN}$$
and TP= True Positives (the number of correct interactions identified), FP= number of pairs incorrectly identified, and FN=number of interactions that were unidentified.  In general, there is a trade-off between precision and recall, as we slide the threshold parameter.

\section{Conclusions}
We have described and detailed  significant problems with the definition and estimation of mutual information, a quantity that is central to information theory. We argue that this issue is of a theoretical nature leading to an inherent arbitrariness in its estimation from a discrete set of points. Moreover, the the errors introduced by the standard forms of partitions of space is explicitly demonstrated with examples .  \\

By formulating a novel adaptive partitioning algorithm for reverse engineering of networks, we further show the impact of these estimation biases in a real-world context where mutual information is applied extensively. The superior performance of the method over standard approaches  on both in-silico and real biological networks is not only an advancement in that field, but also clearly implies the necessity to carefully understand the problems with estimating mutual information. \\

The applications of information theory to various fields have grown enormously in recent decades as the concepts have become central to areas such as physics, communications and signaling, inference theory, multiple biological sciences, pattern recognition and artificial intelligence\cite{Jaynes,Adami,Maas,Per,Tishby,Studholme}.   
We believe that a thorough investigation of the fundamental issues involved in the subject would be immensely beneficial to all of their applications.

\begin{appendices}
\section{Proof of increase in mutual information}
We show that doubling the partition number along one of the directions leads to a non-negative change in the mutual information. Starting with a division of the 2D space into $L \times M$ blocks we bisect each of the $LM$  units along the $X-$axis to create $2L \times M$ cells. The mutual information: 
\begin{equation}
I = \sum_{i=1}^{L} \sum_{j=1}^{M} P_{X,Y} (i,j) \log \frac{P_{X,Y}(i,j)}{P_X(i) P_Y(j)} 
\end{equation}
where $P_{X,Y}(i,j)$ is the fraction of particles lying in the rectangular cell $(i,j)$. After bisection, we have a new probability distribution $P^{2L}(i,j)$ (dropping the double-index subscript) where $i =1,2,\cdots 2L$, such that $P^{2L} (2k-1,j) + P^{2L} (2k,j) = P(k,j)$ for $ k=1,2,\cdots L$. We will show that for every original cell, its contribution to MI is exceeded by the sum of the contributions of the bisected components and thus obtain a stronger version of our required result. 
To prove: 
\begin{equation}
P^{2L} (2k-1,j) \log \frac{P^{2L}(2k-1,j)}{P_X^{2L} (2k-1) P_Y(j)} +  P^{2L} (2k,j) \log \frac{P^{2L} (2k,j)}{P_X^{2L} (2k) P_Y (j)}  \geq P (k,j) \log \frac{P (k,j)}{P_X (k) P_Y (j)}
\end{equation}
which reduces to,
\begin{equation}
P^{2L} (2k-1,j) \log \frac{P^{2L}(2k-1,j)}{P_X^{2L} (2k-1)} +  P^{2L} (2k,j) \log \frac{P^{2L} (2k,j)}{P_X^{2L} (2k)}  \geq P (k,j) \log \frac{P (k,j)}{P_X (k)}.
\end{equation}
Substituting $p_a = P^{2L} (2k-1,j), p_b  = P^{2L} (2k,j),p_1= P_X^{2L} (2k-1), p_2 = P_X^{2L} (2k)$, 
\begin{equation}
p_a \log \frac{p_a}{p_1} + p_b \log \frac{p_b}{p_2} \geq (p_a + p_b) \log \frac{p_a +p_b}{p_1 +p_2}
\end{equation}     
 Straightforward rearrangements of the terms leads to,
 \begin{align}
  \frac{p_a}{p_a +p_b} \log \frac{p_a}{p_a + p_b} &+  \frac{p_b}{p_a +p_b} \log \frac{p_b}{p_a + p_b}  - \frac{p_a}{p_a +p_b} \log \frac{p_1}{p_1 + p_2} \\ \notag &-  \frac{p_b}{p_a +p_b} \log \frac{p_2}{p_1 + p_2} \geq 0 
 \end{align}
 which is equivalent to demonstrating that :
 
 \begin{equation}
 x \log \frac{x}{y}  +(1-x) \log \frac{1-x}{1-y} \geq 0
 \end{equation}
 for $0\leq x,y \leq 1$.  The left hand side is nothing but KL-divergence between two distributions on two element set with probabilities $\{ x,1-x \}$ and $ \{ y,1-y  \}$, which is always positive. 

\section{Modified Algorithm for determining targets}
We used a modification of the general algorithm for the TF-target identification in the {\it E. coli} expression data set.  As the set of transcription factors are known, the binning adjustments can be limited to the targets here and thus our technique is significantly simplified.  As before $w_0 = w_{int}/H$, where $w_{int} = \text{median} \{ \sigma (X) | X \in V  \}$, and $H$ and $K$ represent the minimum and maximum number of bins. \\

{\bf Step 1}: Fix $b_X = (H+K)/2$ \\
{\bf Step 2} : If $b_Y > K$, then reset $b_Y= K$.    \\
{\bf Step 3}: If $b_Y <H$, then $b_Y = H$. \\
{\bf Step 4}: Once the binning is fixed, we proceed to calculate the number of points falling within each rectangle, and the discretized form of mutual information between the two variables. \\
{\bf Step 5}: We normalize the mutual information by dividing by $\min \{ H(X), H(Y) \}$, where $H(X)$ and $H(Y)$ are the entropies of $X$ and $Y$ calculated using the same bin numbers $b_X$ and $b_Y$ respectively.  
\end{appendices}


\begin{thebibliography}{22}

\bibitem{Tegner} Yeung M.K., Tegn\'er J., Collins J.J., {\bf Reverse engineering gene networks using singular value decomposition and robust regression}, {\it Proc Natl Acad Sci USA}, {\bf 99(9)} :6163-6168 (2002).
\bibitem{Hecker} Hecker M.,Lambeck S., Toepfer S.,van Someren  E., Guthke R., {\bf Gene regulatory network inference: Data integration in dynamic models—A review}, BioSystems {\bf 96}  86–103 (2009)
\bibitem{Markow} Markowetz F.,Span R., {\bf Inferring cellular networks – a review}, {\it BMC Bioinformatics } {\bf 8} (Suppl 6):S5 (2007)  
\bibitem{Bansal} Bansal M.,Belcastro V., Alberto Ambesi-Impiombato1, di Bernardo D., {\bf How to infer gene networks from expression profiles}, {\it Molecular Systems Biology }{\bf 3}:78 (2007)
\bibitem{Hendrick}  Hendricks D.M.,  Hendriks M.M. W. B.,Eilers  P. H. C.,Smildeac A.K., Hoefslootac Huub C. J.,{\bf Reverse engineering of metabolic networks, a critical assessment}, {\it Mol. BioSyst.}, {\bf 7}, 511 (2011)  
\bibitem{REVEAL}  Liang S, Fuhrman S and Somogyi R., {\bf Reveal, a general reverse engineering algorithm for inference of genetic network architectures}, {\it Pac Symp Biocomput}, 18 (1998)
\bibitem{Butte} Butte A.J.,Kohane I.S., {\bf Mutual information relevance networks: functional genomic clustering using pairwise entropy measurements.}, {\it Pac Symp Biocomput.},{\bf 5 } 415 (2000)
\bibitem{Reshef}  Reshef D.N., Reshef Y.A., Finucane H.K., Grossman S.R., McVean G., Turnbaugh P.J., Lander E.S., Mitzenmacher M., Sabeti P.C., {\bf Detecting Novel Associations in Large Data Sets}, {\it Science} {\bf 334} (6062): 1518 (2011)
\bibitem{Faith} Faith J.J., Hayete B.,Thaden J.T.,Mogno I., Wierzbowski J., Cottarel G.,Kasif S., Collins J.J., Gardner T.S., { \bf Large-Scale Mapping and Validation of Escherichia coli Transcriptional Regulation from a Compendium of Expression Profiles}, {\it PLoS Biol} {\bf 5(1)}: e8.
\bibitem{INem} Margolin A.A., Wang K., Califano Nemenman I, {\bf Multivariate dependence and genetic networks inference}, {\it IET Syst Biol.} {\bf 4(6) }:428
\bibitem{Shan} Shannon C.E., {\bf The mathematical theory of communication },{\it Bell Syst. Techn. Journal}, {\bf 27}, 379 (1948). 
\bibitem{Lesne} Lesne A.,{\bf Shannon entropy: a rigorous notion at the crossroads between probability, information theory, dynamical systems and statistical physics}, {\it Mathematical Structures in Computer Science}, {\bf 24}, SI 03, 24031 (2014) 
%\bibitem{Steuer} Steuer R., Kurths J., Daub C.O., Weise J., Selbig J., {\bf The mutual information: Detecting and evaluating dependencies between variables}, {\it Bioinformatics}, {\bf 18} (suppl 2), S231-S240 (2002)
%\bibitem{Olsen} Olsen .,Meyer P.E.,Bontempi G., {\bf On the Impact of Entropy Estimation on Transcriptional Regulatory Network Inference Based on Mutual Information}, {\it EURASIP Journal on Bioinformatics and Systems Biology} {\bf 2009}: 308959 (2008)
\bibitem{Simoes} de Matos Simoes R., Emmert-Streib F., {\bf Influence of Statistical Estimators of Mutual Information and Data Heterogeneity on the Inference of Gene Regulatory Networks} {\it PLoS ONE} {\bf 6(12)}: e29279 (2011)
%\bibitem{Hausser} Hausser J., Strimmer K., {\bf Entropy Inference and the James-Stein Estimator, with Application to Nonlinear Gene Association Networks} {\it The Journal of Machine Learning Research} {\bf  10},  1469-1484 
%\bibitem{Altay}  Altay G.,Emmert-Streib F., {\bf Inferring the conservative causal core of gene regulatory networks}, {\it BMC Systems Biology}, {\bf 4}:132 (2010)
%\bibitem{MRNET} Meyer PE, Kontos K, Lafitte F, Bontempi G., {\bf Information-theoretic inference of large transcriptional regulatory networks}, {\it EURASIP J Bioinform Syst Biol.}, {\bf 2007}(1): 79879.
\bibitem{KBE} Moon Y.I., Rajagopalan B., Lall U.,{\bf Estimation of mutual information using kernel density estimators}, {\it Phys. Rev. E} {\bf 52}, 2318 (1995)
\bibitem{ARACNE} Margolin A.A., Nemenman I., Basso K., Wiggins C., Stolovitzky G., Dalla Favera R., Califano A., {\bf ARACNE: an algorithm for the reconstruction of gene regulatory networks in a mammalian cellular context},  {\it BMC Bioinformatics} {\bf 7} (Suppl 1):S7
\bibitem{Knn} Kraskov A.,Stögbauer H., Grassberger P., {\bf Estimating Mutual Information}, {\it Phys. Rev. E} {\bf 69}, 066138 (2004)
\bibitem{Chow} Chow CK, Liu CN, {\bf Approximating discrete probability distributions with dependence trees},{\it IEEE Trans. Inf. Thy.},{\bf 14(3)}, 462 (1968)
\bibitem{Marbach} Marbach {\it et al}, {\bf Wisdom of crowds for robust gene network inference}, {\it Nature Methods}, {\bf 9}, 796 (2012)
%\bibitem{Fraser-Swinney} Fraser A.M., Swinney H., {\bf Independent coordinates for strange attractors from mutual information}, {\it  Phys. Rev. A} {\bf 33}, 1134 (1986)
%\bibitem{Darbellay} Darbellay G.A., Vajda I., {\bf Estimation of the Information by an Adaptive Partitioning of the Observation Space}, {\it IEEE Transcations on Information Theory} {\bf 45} (1315) (1999)
\bibitem{Jaynes} Jaynes E.T., {\bf Information Theory and Statistical Mechanics}, {\it Phys. Rev.}, {\bf 106}, 620 (1957)
\bibitem{Adami}  Adami C., {\bf Information theory in molecular biology}, {\it Physics of Life Reviews}, {\bf 1(1)}, 3 (2004)
\bibitem{Maas}  Maasoumi E., {\bf A compendium to information theory in economics and econometrics}, {\bf 12 (2)}, 137 (1993)
\bibitem{Per} Pereira F., {\bf Formal grammar and information theory: together again? }, {\it Phil. Trans. R. Soc. Lond. A }, {\bf  358}, 1239 (2000) 
\bibitem{Tishby}  Tishby N., Pereira F.C., Bialek W., {\bf The Information Bottleneck method}, {\it The 37th annual Allerton Conference on Communication, Control, and Computing}, 368–377 (1999)
\bibitem{Studholme}  Studholme C., Hill D.L.G., Hawkes D.J., {\bf An overlap invariant entropy measure of 3D medical image alignment} {\it Pattern Recognition} {\bf 32(1)}  71–86 (1999)
\end{thebibliography}
\end{document}